\documentclass[prb,twocolumn,aps,noshowpacs]{revtex4}
\usepackage{graphicx,epsfig}
\usepackage{bm}
\usepackage{amssymb} 
\usepackage{bbold}

\begin{document}

\title{Comparative analysis of magnetic resonance in the polaron pair recombination and the triplet
exciton-polaron quenching models}

\author{ V. V. Mkhitaryan,$^{1,*}$ D. Danilovi\`{c},$^1$ C. Hippola,$^1$  M. E. Raikh,$^2$ and J. Shinar$^1$}

\affiliation{$^1$Ames Laboratory, Iowa State University, Ames,
Iowa 50011, USA}

\altaffiliation[Corresponding author:] { vmkhitar@ameslab.gov}

\affiliation{$^2$Department of Physics and Astronomy, University
of Utah, Salt Lake City, UT 84112}

\begin{abstract}

We present a comparative theoretical study of magnetic resonance
within the polaron pair recombination (PPR) and the triplet
exciton-polaron quenching (TPQ) models. Both models have been
invoked to interpret the photoluminescence detected magnetic
resonance (PLDMR) results in $\pi$-conjugated materials and
devices.
We show that resonance lineshapes calculated within the two models
differ dramatically in several regards. First, in the PPR model,
the lineshape exhibits unusual behavior upon increasing the
microwave power: it evolves from fully positive at weak power to
fully negative at strong power. In contrast, in the TPQ model, the
PLDMR is completely positive, showing a monotonic saturation.
Second, the two models predict different dependencies of the
resonance signal on the photoexcitation power, $P_L$. At low
$P_L$, the resonance amplitude $\Delta I/I$ is $\propto P_L$
within the PPR model, while it is $\propto P_L^2$ crossing over to
$P_L^3$ within the TPQ model. On the physical level, the
differences stem from different underlying spin dynamics. Most
prominently, a negative resonance within the PPR model has its
origin in the microwave-induced spin-Dicke effect, leading to the
resonant quenching of photoluminescence. The spin-Dicke effect
results from the spin-selective recombination, leading to a highly
correlated precession of the on-resonance pair-partners under the
strong microwave power. This effect is not relevant for TPQ
mechanism, where the strong zero-field splitting renders the
majority of triplets off-resonance. On the technical level, the
analytical evaluation of the lineshapes for the two models is
enabled by the fact that these shapes can be expressed via the
eigenvalues of a complex Hamiltonian. This bypasses the necessity
of solving the much larger complex linear system of the stochastic
Liouville equations. Our findings pave the way towards a reliable
discrimination between the two mechanisms via cw PLDMR.

\end{abstract}

\maketitle

\section{Introduction}

Over the years, optically detected magnetic resonance (ODMR)
has  proven to be a powerful tool for the study of spin-dependent
recombination and dissociation processes, both in inorganic
\cite{Cavenett81, Street82, Depinna82} and in organic
\cite{ShinarLPR12} semiconductors. High sensitivity, exceeding the
sensitivity of conventional electron spin resonance by about six
orders of magnitude, renders ODMR the tool of choice when it comes
to $\pi$-conjugated polymers, \cite{ShinarLPR12} where the density
of spin carriers is small. Photoluminescence detected magnetic
resonance (PLDMR), being  a subset of ODMR, has an advantage as it
provides the most straightforward probe of the radiative singlet
exciton population and quantum yield of the
material.\cite{ShinarLPR12, Wohlgenannt01, ShinarPRL05,
VardenyPRL07} Besides, this method is suitable for probing the
bulk of a photoluminescent material without the necessity of
device fabrication.

Two different models have been employed to explain PLDMR results
in $\pi$-conjugated materials. The double modulation PLDMR
experiment \cite{ShinarPRL05} advocated the quenching model based
on the spin-dependent reaction between triplet excitons and
polarons (TPQ). On the other hand,
the experimental study of the frequency dependence of the in-phase
component of PLDMR \cite{VardenyPRL07} employed the polaron pair
recombination model (PPR). Subsequent publications invoked both
the TPQ model \cite{ShinarPRB05, ShinarPRB15} and the PPR model,
\cite{VardenyPRB08} for the interpretation of results obtained for
the same material,
polymer MEH-PPV.

In order to distinguish between the two models, pulsed PLDMR
experiment were conducted, \cite{BoehmeJacs11} in which Rabi beats
of PLDMR in
MEH-PPV and its deuterated variant were
explored. The results appear to reveal the fingerprints of both
the PPR and the TPQ mechanisms. Hence, for conclusive
discrimination, additional continuous wave PLDMR measurements
revealing the nature of the underlying spin-dependent processes
are desirable. Equally, theoretical predictions of the differences
in the PLDMR within the two models are highly desirable. That is
the goal of the present paper. To achieve this goal, we employ the
stochastic Liouville equations for the density matrix to calculate
analytically the resonance lineshapes and saturation within the
PPR and TPQ models. We show how the difference in the underlying
spin dynamics translates into very different dependencies of the
PLDMR on the optical excitation intensity. Also, within the PPR
model, the lineshape is predicted to be very peculiar, with a peak
precisely at the resonant frequency evolving into a minimum at
higher microwave power.

Our results on the dynamics of the spin pairs within the PPR model
agree with the predictions based on the analysis of eigenmodes for
the calculation of transport, \cite{RRSlow, RRResonant} and with a
direct analytical solution of the Liouville equations.
\cite{Lvov82, Barabanov96}

We consider the regime relevant to fluorescent $\pi$-conjugated
polymers (very close electron and hole polaron $g$- factors,
relatively strong hyperfine interaction, relatively slow PPR from
the singlet PP state, relatively slow annihilation of triplet
excitons from doublet triplet-polaron state, relatively long spin
coherence times, weak exchange, and weak dipolar interaction
between PPs or triplet exciton-polaron spins, etc.)

Our analytical results can be directly generalized to include a
broader class of ODMR techniques, as well as other detection
methods, e.g., electrical, reaction yield, and capacitance
measurements. In this connection, notice the similarity between
our results for the PPR model and those observed in recent
transport \cite{Waters15} and dielectric polarizability
\cite{Bayliss15} studies.

The established substantial differences between the predictions of
the PPR and TPQ models can enable the differentiation of the two
mechanisms in interpretation of continuous wave PLDMR results.


\section{The polaron pair recombination model}

\subsection{Qualitative picture}

\begin{figure}[t]
\centerline{\includegraphics[width=60mm,angle=0,clip]{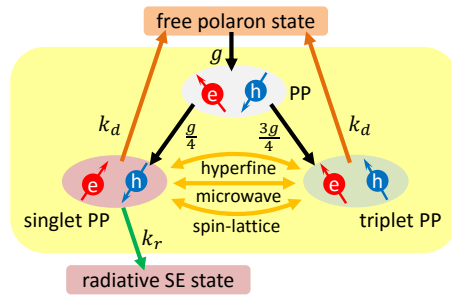}}
\caption{(Color online) Schematics of the processes underlying the
PPR model. The black arrows indicate the PP generation; out of $g$
PPs per
second, $\frac g4$ are singlets and $\frac{3g}4$ -- triplets. The
brown and green arrows respectively indicate the dissociation and
recombination, Eq. (\ref{Rgr}). The orange arrows represent the
singlet-triplet beating, induced by the hyperfine coupling and
resonance microwave, Eq. (\ref{Hamil}), and the spin-lattice
relaxation, Eq. (\ref{Rsl}). The yellow background outlines the
components of the stochastic Liouville Equation (\ref{sLe}).
} \label{PPRschem}
\end{figure}

The PPR model is illustrated in Fig. \ref{PPRschem}. The processes
involved are the generation of weakly coupled PPs at rate $g$,
their dissociation with at rate $k_d$, and recombination from the
singlet pair state at rate $k_r$. \cite{Boehme03} The latter
process constitutes the reaction,
\begin{equation}\label{PPreact}
\text{P}_e + \text{P}_h \rightarrow \text{S},
\end{equation}
between the electron and hole polarons, P$_e$ and P$_h$
respectively, yielding a singlet exciton, S. Thus, the
spin-selective recombination is incorporated as the restriction
that Eq. (\ref{PPreact}) can occur only for singlet PPs, i.e., for
triplet PPs it is forbidden.

In an applied static magnetic field,
$\mathbf{B}_0=B_0\hat{\mathbf{z}}$, the electron- and hole-polaron
spins occupy the Zeeman levels, $\pm \frac12\hbar\gamma B_0$,
where $\gamma$ is the polaron gyromagnetic ratio (we assume equal
gyromagnetic ratios for the electron- and hole-polarons). The
resonant microwaves couple the Zeeman levels of individual spins
and, correspondingly, the triplet PP levels. On the other hand,
random hyperfine fields created by the nuclei (almost entirely
hydrogen protons) at electron- and hole-polaron sites induce
interconversion between the singlet and triplet PP levels.
Characteristic magnitudes of these hyperfine fields $b_{\text{hf},
e}$ and $b_{\text{hf}, h}$ are different in general and define two
distinct hyperfine frequencies, $\omega_{\text{hf}, \mu}=\gamma
b_{\text{hf}, \mu}$, $\mu = e$, $h$.

If the pair spins are uncorrelated the populations of individual
Zeeman levels in the microwave drive field $\mathbf{B}_1(t) =
2B_1\cos(\omega t) \hat{\mathbf{x}}$ oscillate with Rabi
frequencies
\begin{equation}\label{Rabis}
\Omega_\mu = \sqrt{\omega_\mu^2 +\omega_1^2}, \quad \mu = e,\,h,
\end{equation}
where $\omega_1= \gamma B_1$ is the microwave drive amplitude,
and
\begin{equation}\label{oms}
\omega_\mu = \gamma b_{z,\mu} +\delta,\quad \delta = \gamma B_0
-\omega,
\end{equation}
are the polaron Larmor frequencies in the rotating frame and the
detuning frequency, respectively. The $z$-components of the random
hyperfine fields $b_{z,\mu}$ follow a Gaussian distribution,
entailing a Gaussian distribution of Larmor frequencies:
\begin{equation}\label{GaussDis}
\mathcal{N}(\omega_\mu) = \frac1{\sqrt{2\pi}\omega_{\text{hf},\mu}
} e^{-(\omega_\mu-\delta)^2/2\omega_{\text{hf}, \mu}^2},\quad \mu
= e, h.
\end{equation}

The most important physics of the PPR model is that the
spin-selective recombination correlates the dynamics of each of
the the spins in the pair. Indeed, if the recombination rate was
the same for all four spin-pair states, then the Rabi beatings of
the level populations would not affect the luminescence, and
therefor no PLDMR would be detectable.

The essence of PLDMR technique is that the intensity of
recombination shown in Fig. \ref{PPRschem} exhibits a resonance as
a function of $\delta$, which becomes progressively pronounced as
the microwave field amplitude $\omega_1$ exceeds
$\omega_{\text{hf}, \mu}$. The PLDMR amplitude is directly related
to the singlet exciton density $n_{\text{S}}$. Therefore, the
evaluation of resonance lineshapes reduces to finding
$n_{\text{S}}$ versus $\delta$,  $\omega_1$, and
$\omega_{\text{hf}, \mu}$. One can naturally distinguish two
regimes: weak drive, $\omega_1\ll \omega_{\text{hf}, \mu}$, and
strong drive, $\omega_1\gg \omega_{\text{hf}, \mu}$. It might seem
that, at weak drive Rabi oscillations do not occur. However, as
the hyperfine fields are random, some spins will be at resonance.
Their fraction can be estimated as \cite{Boehme03, McCameyPRL10,
GR13} $\sim\omega_1/\omega_{\text{hf}, \mu}$. Our main result for
weak drive is that these pairs dominate the resonance line shape,
leading to the linear dependence of the resonance amplitude on
$\omega_1$. This conclusion contrasts with the results obtained
from simple rate equations \cite{VardenyPRL07} and from other
studies of PP spin dynamics and recombination that exclude
averaging over local hyperfine fields. \cite{Eickelkamp, Lvov82}

In the strong drive regime, the physics underlying the resonance
line shape is different. In this regime, not only are the four
conventional spin-pair states not eigenstates, but actually
$\bigl( |T_{+1} \rangle -|T_{-1} \rangle \bigr )/\sqrt{2}$ is
close to an eigenstate, and it is decoupled from $|S\rangle$. This
means it is a long-lived state. We will see that this decoupling
is a consequence of the spin Dicke effect. \cite{RRSlow,
RRResonant} It manifests itself as a minimum in the resonance line
shape at zero detuning, which gradually takes over as the
microwave drive increases, turning the resonance to fully
negative.

More formally, under steady state conditions, the
photoluminescence intensity, $\mathcal{I}$, is proportional to the
steady state singlet density, $\tilde{n}_{\text{S}}$. The latter
is found from the rate equation,
\begin{equation}\label{PPnS}
\partial_t n_{\text{S}} = G_S - R_S n_{\text{S}}
+ \alpha(\delta,\omega_1),
\end{equation}
where $G_S$ is the photoexcitation rate of singlet excitons, $R_S$
is their decay rate, and  $\alpha(\delta, \omega_1)$ is the
rate of singlet exciton generation due to the PPR, Eq.
(\ref{PPreact}), rendering the PLDMR within the PPR model. The
normalized PLDMR is then given by
\begin{equation}\label{nrmI}
\frac{\Delta\mathcal{I}(\delta,\omega_1)}{\mathcal{I}(0)} =
\frac{\tilde{n}_{\text{S}}(\delta, \omega_1)
-\tilde{n}_{\text{S}}(0)}{ \tilde{n}_{\text{S}}(0)} =
\frac{\alpha(\delta, \omega_1)-\alpha(0)}{ G_S},
\end{equation}
where the relation, $G_S\gg \alpha$, common for many systems,
\cite{ShinarLPR12} is used in the last equality, and zero
arguments correspond to $\omega_1=0$, implying the absence of
microwave drive;
$\alpha(\delta, \omega_1)$ is governed by the spin dynamics of
polaron pairs, to which we turn next.

\subsection{Spin dynamics of weakly coupled electron-hole pair ensemble}

The spin dynamics of a PP ensemble is analyzed by solving the
stochastic Liouville equation for the spin density matrix $\rho$,
\begin{equation}\label{sLe}
\frac {d \rho}{dt}= i[\rho,H] + \frac g4 \mathbb{1} +
\mathcal{R}_{\text{dr}} \{ \rho \} + \mathcal{R}_{\text{sl}} \{
\rho \},
\end{equation}
where the first term describes the spin dynamics due to the
magnetic interactions governed by the spin Hamiltonian $H$, $g$ is
the PP generation rate, $\mathbb{1}$ is the identity operator,
$\mathcal{R}_{\text{dr}}$ represents the pair dissociation and
recombination, and $\mathcal{R}_{\text{sl}}$ -- the spin-lattice
relaxation processes.

For simplicity, we neglect the spin exchange and dipolar
interactions (generalization for the non-zero spin exchange and
dipolar interactions will be discussed later). In the rotating
frame, the spin Hamiltonian is given by
\begin{equation}\label{Hamil}
H = \omega_e S_e^z +\omega_h S_h^z +\omega_1(S_e^x +S_h^x),
\end{equation}
where $\omega_e$ and $\omega_h$ are local electron and hole
detunings, see Eq. (\ref{oms}). They are different due to the
different on-site hyperfine fields. $\mathbf{S}_{e,h}$ stand for
the electron- and hole-polaron spin operators (we set $\hbar =
1$).

Conventionally, the spin-dependent recombination processes are
described within the singlet-triplet basis of PP. We assume that
the pair dissociation occurs at the equal rate $k_d$ from all spin
states. In terms of the matrix elements we have
\begin{equation}\label{Rgr}
\mathcal{R}_{\text{dr}}\{\rho\}_{\alpha \beta}= - k_d\rho_{\alpha
\beta} -\frac{k_r}2(\delta_{\alpha S} +\delta_{S \beta})
\rho_{\alpha \beta},
\end{equation}
where $\alpha, \beta = +1$, $-1$, $0$, and $S$ enumerate the
singlet and triplet spin states $|T_{+1}\rangle$,
$|T_{-1}\rangle$, $|T_0\rangle$, and $|S\rangle$, respectively.

For the spin-lattice relaxation we take the form, \cite{Lvov82}
\begin{equation}\label{Rsl}
\mathcal{R}_{\text{sl}}\{\rho\}_{\alpha \beta}= -\bigl(
1/T_{\text{sl}} \bigr )\bigl[\rho_{\alpha \beta} -\delta_{\alpha
\beta} \text{tr}(\rho/4) \bigr].
\end{equation}
This relaxation tends to equalize the state populations, with the
rate $1/T_{\text{sl}}$.

As an important step, we introduce the complex Hamiltonian,
\begin{equation}\label{compHam}
\mathcal{H} = H-i(w_d/2)\mathbb{1} -i(k_r/2)\Pi_S,
\end{equation}
where $w_d=k_d+1/T_{\text{sl}}$ and $\Pi_S =|S\rangle\langle S|$
is the projection operator onto the singlet state. In terms of the
complex Hamiltonian, Eq. (\ref{sLe}) for the density matrix takes
the form,
\begin{equation}\label{rewsLe}
\frac {d \rho}{dt}= i\bigl(\rho\mathcal{H}^* -\mathcal{H}\rho
\bigr) + \frac 14 \bigl( g + T_{\text{sl}}^{-1} \text{tr}\rho
\bigr) \mathbb{1}.
\end{equation}
The observable quantities are described by the steady state
density matrix, $\tilde{\rho}$, satisfying
\begin{equation}\label{sdm}
i\bigl(\tilde{\rho}\mathcal{H}^* -\mathcal{H}\tilde{\rho} \bigr) +
\frac 14 \bigl( g + T_{\text{sl}}^{-1} \text{tr}\tilde{\rho}
\bigr) \mathbb{1}=0.
\end{equation}
We write the formal solution of Eq. (\ref{sdm}) as
\begin{equation}\label{ssrho}
\tilde{\rho}= \frac14\bigl( g + T_{\text{sl}}^{-1}
\text{tr}\tilde{\rho} \bigr) U,\quad U= \int_0^\infty \!\! dt\,
e^{-i\mathcal{H}t}e^{i\mathcal{H}^*t}.
\end{equation}
Thus, the matrix structure of $\tilde{\rho}$ is posed by $U$.
Another useful relation is found by taking the trace of the right
hand side of Eq. (\ref{sdm}):
\begin{equation}\label{tr}
g -k_d\text{tr}\tilde{\rho} -k_r \tilde{\rho}_{SS} =0,
\end{equation}
where $\tilde{\rho}_{SS} = \langle S| \tilde{\rho} |S \rangle$ is
the singlet polaron pair population. Equation (\ref{tr}) is the
balance equation between the generation of PPs and their
destruction, taking place from the triplet and singlet states with
the rates $k_d$ and $k_d +k_r$, respectively. From
Eqs. (\ref{ssrho}) and (\ref{tr})  we find:
\begin{equation}\label{Lviatr}
k_r \tilde{\rho}_{SS} = g L(\delta, \omega_1), \quad L= \frac{1 -
(w_d/4)\text{tr}U(\delta, \omega_1)} { 1 -
(1/4T_{\text{sl}})\text{tr} U(\delta, \omega_1)}.
\end{equation}
As will be seen shortly, Eqs. (\ref{ssrho}) and (\ref{Lviatr}) are
sufficient for the calculation of $\alpha(\delta,\omega_1)$, which
is the resonance lineshape. Most importantly, we will need only
the eigenvalues of $4 \times 4$ complex Hamiltonian $\mathcal{H}$.
This bypasses the necessity of solving effectively $10 \times 10$
complex system of linear equations (\ref{sdm}).

The calculation of resonance lineshape, $\alpha(\delta,
\omega_1)$, involves also the averaging over the Gaussian
distribution of hyperfine Larmor frequencies,
Eq.~(\ref{GaussDis}):
\begin{equation}\label{AlpviaL}
\alpha(\delta, \omega_1) =  k_r \langle\tilde{\rho}_{SS}
\rangle_{\text{hf}} = g \langle L(\delta, \omega_1)
\rangle_{\text{hf}}.
\end{equation}
Furthermore, for the PP generation rate one has
$g\propto\tilde{n}_{\text{P}}^2$, where $\tilde{n}_{\text{P}}$ is
the steady state density of polarons.
Thus, from Eqs. (\ref{nrmI}) and (\ref{AlpviaL}) we write:
\begin{equation}\label{NormdL}
\frac{\Delta\mathcal{I}}{\mathcal{I}} =
\frac{\lambda_{P}\tilde{n}_{\text{P}}^2}{ G_S} \mathcal{L}
(\delta, \omega_1),\quad \mathcal{L} = \bigl \langle L(\delta,
\omega_1) - L(0) \bigr \rangle_{\text{hf}},
\end{equation}
where the constant, $\lambda_{P}$, is determined by the PP
formation cross section, proportional to the polaron mobility.

\begin{figure}[t]
\vspace{-0.4cm}
\centerline{\includegraphics[width=95mm,angle=0,clip]{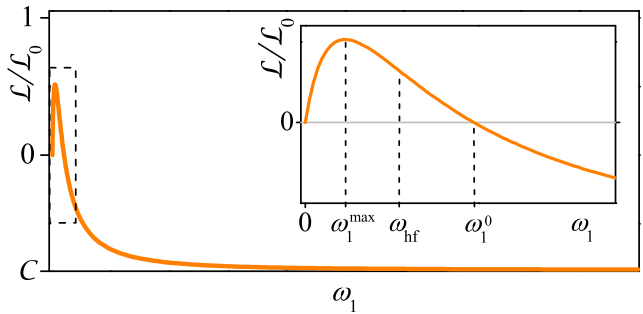}}
\vspace{-0.5cm} \caption{(Color online) Sketch of the function,
$\mathcal{L} (\omega_1)$, defining the PLDMR amplitude at zero
detuning, $\delta=0$, via Eq.~(\ref{NormdL}), in units of
$\mathcal{L}_0$, Eq. (\ref{L0anda}). The constant $\mathcal{C}$,
given by Eq. (\ref{bK}), is the saturation value of $\mathcal{L}
(\omega_1)/\mathcal{L}_0$. The inset zooms into the region
indicated in the main panel with a dashed rectangle. }
\label{resdrive}
\end{figure}

From now on we focus on the regime of weak recombination,
$k_r\ll\omega_{\text{hf}}$. Still, before going into the details
of analytical calculation, we outline in Fig. \ref{resdrive} the
result obtained by solving Eq. (\ref{sdm}) and performing the
averaging in Eq. (\ref{AlpviaL}) numerically. The plot in Fig.
\ref{resdrive} shows a steep increase at weak $\omega_1\ll
\omega_{\text{hf} }$, a maximum followed by a moderate decrease at
$\omega_1\gtrsim\omega_{\text{hf}}$, and a very slow decrease to
negative values with saturation at the strongest drives. This
picture appears to be quite general for a wide range of model
parameters. In addition, in the limits of weak and strong drive
the curve can be described analytically. This is accomplished in
the next subsection.

\subsection{Perturbation with respect to small $k_r$}

In the limit of slow recombination, $k_r\ll\omega_{\text{hf}}$,
the perturbative approach applies. The unperturbed eigenstates of
$\mathcal{H}$ are the eigenvectors of the Hamiltonian
(\ref{Hamil}):
\begin{equation}\label{Hunp}
H|\varphi_\alpha \rangle =\epsilon_\alpha|\varphi_\alpha\rangle,
\quad \alpha =1,..., 4.
\end{equation}
In the absence of recombination the pair partners are independent,
so that the eigenvalues are given by
\begin{equation}\label{e0}
\epsilon_1= -\epsilon_4= \frac12(\Omega_e+\Omega_h),\,\,\,
\epsilon_2= -\epsilon_3= \frac12(\Omega_e-\Omega_h),
\end{equation}
where $\Omega_{e,h}$ are defined by Eq. (\ref{Rabis}).

In the presence of recombination, the eigenvectors are perturbed
by the operator $V=-i(k_r/2) \Pi_S$, which is responsible for this
process. The matrix form of this operator, $V_{\alpha \beta}=
\langle \varphi_\alpha | V| \varphi_\beta\rangle$, is found in
Appendix \ref{AppPP}. It is conveniently parameterized by the
angles,
\begin{equation}\label{angles}
\tan 2\phi_\mu = \frac{\omega_1}{\omega_\mu}, \quad \mu=e,h.
\end{equation}
The explicit form of the matrix reads:
\begin{equation}\label{Vab}
V=-i\frac{k_r}4 \frac 1{1+\xi^2} \left(\!\begin{array}{cccc}
\xi^2&\xi &-\xi &\xi^2\\
\xi &1 &-1 &\xi\\
-\xi & -1 &1 &-\xi\\
\xi^2 &\xi &-\xi &\xi^2
\end{array}\!\right)\! ,
\end{equation}
where
\begin{equation}\label{tanphi}
\xi = \tan (\phi_{eh}), \quad \phi_{eh}=\phi_e - \phi_h.
\end{equation}
The leading recombination-induced corrections to the eigenvalues
Eq. (\ref{e0}) are given by the diagonal elements;
\begin{equation}\label{leadcorr}
\epsilon_{1,4}^{(1)}= -i\frac{k_r}4\sin^2\phi_{eh}, \quad
\epsilon_{2,3}^{(1)} = -i\frac{k_r}4\cos^2\phi_{eh}.
\end{equation}

According to the standard perturbation theory, \cite{LandLif} the
eigenstates of $\mathcal{H}$ are close to
$|\varphi_\alpha\rangle$, when $|\epsilon_\alpha
-\epsilon_\beta|\gg k_r$ for $\alpha \neq \beta$. Here we make a
crucial observation that for certain pairs for which $\Omega_e$
and $\Omega_h$ are anomalously close, this condition is violated.
Such a ``softening'' of modes manifests the degeneracy in the
perturbation theory. As a result, the eigenstates of $\mathcal{H}$
strongly deviate from $|\varphi_2\rangle$ and $|\varphi_3\rangle$,
and are determined by the small $V$.

The condition of softening is progressively satisfied as the drive
increases. This is because $|\Omega_e-\Omega_h|\approx
|\omega_e^2-\omega_h^2|/2\omega_1$ decreases with drive. As a
result, $|\varphi_2\rangle$ and $|\varphi_3\rangle$ are close to
\begin{equation}\label{TTS}
\frac12\bigl(|T_+\rangle - |T_-\rangle \pm \sqrt{2}
|S\rangle\bigr),
\end{equation}
whereas the corresponding eigenstates of $\mathcal{H}$ are close
to
\begin{equation}\label{TandS}
\frac1{\sqrt{2}}\bigr(|T_+\rangle - |T_-\rangle \bigl)\quad
\text{and}\quad |S\rangle.
\end{equation}
This in turn suppresses the overall recombination. It is important
to emphasize that the strong modification of eigenstates and the
entailing lifetime anomaly is the consequence of the back-action
of recombination on the quantum dynamics of PP spins. As pointed
out in Refs. \onlinecite{RRSlow, RRResonant}, there is a close
analogy between the long living states and the subradiant modes in
the Dicke effect. \cite{Dicke} In previous studies of
spin-dependent recombination this back-action is neglected (see,
e.g., Ref. \onlinecite{Flatte}).

The region of strong drive, where $|\Omega_e-\Omega_h|\lesssim
k_r$, is difficult to access because of the degeneracy. The
difficulty is circumvented in the following way. Neglecting all
the off-diagonal elements of Eq. (\ref{Vab}), except for $V_{23}$
and $V_{32}$, induces an error in eigenstates and eigenvalues only
of the order of $k_r/\Omega_\mu$ and $k_r^2/\Omega_\mu$,
respectively, whereas the result for $\text{tr}\, U$ remains
correct to the leading order (this is
analogous to the secular approximation widely used in the theory
of magnetic resonance \cite{Slichter}). Therefore we proceed by
replacing $V=-i(k_r/2) \Pi_S$ in Eq. (\ref{compHam}) with
\begin{equation}\label{tildeV}
\tilde{V} = -i\frac{k_r}4 \frac 1{1+\xi^2}
\left(\!\begin{array}{cccc}
\xi^2&0 &0 &0\\
0 &1 &-1 &0\\
0 & -1 &1 &0\\
0 &0 &0 &\xi^2
\end{array}\!\right)\! .
\end{equation}
This
replacement retains all the eigenvalues and eigenvectors of
$\mathcal{H}$ to the leading order, and allows the direct
evaluation of the operator $U$ from Eq. (\ref{ssrho}). We find:
\begin{eqnarray}\label{trUans}
&&\text{tr}\, U(\delta, \omega_1) = \frac{4(4w_d+k_r)}{4w_d^2 +
2w_dk_r
+(k_r^2/4)\sin^2(2\phi_{eh})}\\
&& + \frac{k_r^2\cos^4(\phi_{eh}) } {\bigl[k_r\cos^2(\phi_{eh})
+2w_d\bigr]\bigl[4 \epsilon_2^2 +w_d(k_r\cos^2(\phi_{eh})
+w_d)\bigr]} \nonumber
\end{eqnarray}
(as shown in Appendix \ref{AppPP}, the replacement of $V$ by
$\tilde{V}$, amounts to $\sim(k_r/\Omega_\mu)^2$ order terms, so
that Eq. (\ref{trUans}) is highly accurate). Notably, the
$\delta$- and $\omega_1$- dependence of $\text{tr}\, U$ enters in
Eq. (\ref{trUans}) via the angles, $\phi_{eh}$, and the energy,
$\epsilon_2$. Analytical expression Eq. (\ref{trUans}) is the main
result of this Section. We emphasize again that it is derived
without solving the $10\times10$ equation (\ref{sdm}).

The microscopic origin of the two terms in Eq. (\ref{trUans}) is
easy to trace back. The first term comes from the diagonals of $V$
and describes
the interplay of spin dynamics and recombination far from the
degeneracy. This term is dominant at weak and moderate drive,
$\omega_1\lesssim\omega_{\text{hf}}$. The second term originates
from the off-diagonal elements, $V_{23}$ and $V_{32}$, and becomes
important with the onset of degeneracy. It quantifies the
microwave-induced Dicke effect, prevailing at strong drive,
$\omega_1> \omega_{\text{hf}}$.

The first term in Eq.~(\ref{trUans}) is monotonically decreasing
function of $\sin^22\phi_{eh}$. At the same time, the second term
in Eq.~(\ref{trUans}) is monotonically increasing function of
$\cos^2\phi_{eh}$. This observation yields the estimate,
\begin{equation}\label{ineq}
\frac{16}{4w_d+k_r}\leq \text{tr}\, U \leq \frac{4w_d+3k_r}
{w_d(w_d+k_r)},
\end{equation}
for the upper and lower bounds of $\text{tr}\, U$. For $k_r\ll
w_d$, the left and right sides of Eq. (\ref{ineq}) are both close
to $4/w_d$, while they are quite different in the opposite limit,
$k_r\gg w_d$. This means that, in the first case, magnetic
resonance can induce only a weak relative variations of
$\text{tr}\, U$, and therefore of $\alpha$, whereas a considerable
relative change in $\alpha$ is possible in the latter limit.

\subsection{Averaging over the random hyperfine fields for
slow spin-lattice relaxation}


We defer the discussion of finite spin relaxation to the end of
this Section, and proceed with the case of long coherence time,
$T_{\text{sl}} \gg k_d^{-1}, k_r^{-1}$, From Eq. (\ref{ineq}) it
follows that in this case $(1/4T_{\text{sl}})\,\text{tr}U\ll 1$,
so that the denominator of Eq.~(\ref{Lviatr}) can be treated
perturbatively,
yielding
\begin{equation}\label{simpL}
L= 1- (k_d/4)\text{tr}U.
\end{equation}
Thus, finding the hyperfine average, $\langle L \rangle
_{\text{hf}}$, reduces to averaging of Eq. (\ref{trUans}) over the
Gaussian distribution of Larmor frequencies:
\begin{equation}\label{avtr}
\langle \text{tr}U \rangle_{\text{hf}} = \int d\omega_e d\omega_h
\mathcal{N} (\omega_e) \mathcal{N} (\omega_h)\, \text{tr}U(\delta,
\omega_1)
\end{equation}
(for simplicity we assume that the mean square deviations of the
Gaussian distributions are the same; $\omega_{\text{hf}, e} =
\omega_{\text{hf}, h}= \omega_{\text{hf}}$, unless it is stated
otherwise).

\subsubsection{Zero detuning}

For zero detuning, $\delta=0$, the random variables $x = (\omega_e
+ \omega_h) /2\omega_1$ and $y = (\omega_e - \omega_h) /2\omega_1$
have the same Gaussian distribution,
\begin{equation}\label{Gaussxy}
\mathcal{P} (x)= \frac 1 { \sqrt{\pi} \beta_0}\exp(-x^2
/\beta_0^2), \quad \beta_0= \frac{\omega_{\text{hf}}}{\omega_1},
\end{equation}
Relevant quantities entering in Eq. (\ref{trUans}) are given by
\begin{equation}\label{sinf}
\sin^2(2\phi_{eh})= \frac{4y^2}{(1+x^2-y^2)^2+4y^2},
\end{equation}
and
\begin{equation}\label{eps2}
\varepsilon_2= \frac{\omega_1}2 \left(\sqrt{1+(x+y)^2} -
\sqrt{1+(x-y)^2}\right).
\end{equation}
Below, the averaging is performed analytically, in the limits of
weak and strong drive.

\subsubsection{Weak resonant drive, $\omega_1\ll
\omega_{\text{hf}}$}

In the limit of weak drive the second term of Eq.~(\ref{trUans})
is negligible, because the PP realizations with
$4\varepsilon_2^2\lesssim w_dk_r$, for which this term is
appreciable, have the probability $\sim \sqrt{w_dk_r}/
\omega_{\text{hf}} \ll1$. Therefore, in this limit we neglect the
second term of Eq.~(\ref{trUans}).
For typical pairs under a weak resonant microwave one has $|x|,
|y| \gg 1$, so the approximate relation,
\begin{equation}\label{weaksin}
\sin^2(2\phi_{eh})\approx \frac{4y^2}{(x^2-y^2)^2+4y^2},
\end{equation}
can be used with the first term of Eq. (\ref{trUans}), leading to
\begin{equation}\label{dLweak}
\mathcal{L} (\omega_1) = \mathcal{L}_0 \!\int\! dx dy \mathcal{P}
(x) \mathcal{P} (y) \frac{y^2}{a^2(x^2-y^2)^2 +y^2},
\end{equation}
where
\begin{equation}\label{L0anda}
\mathcal{L}_0=\! \frac{k_dk_r^2}{2w_d( 2w_d+k_r)(4w_d+k_r)},
\,\,\, a=\! \frac{\sqrt{2w_d(2w_d+k_r)}}{4w_d+k_r}.
\end{equation}
For $a\beta_0>1$ ($\omega_1<a\omega_{\text{hf}}$) the integral
(\ref{dLweak}) further simplifies, as in this case it is dominated
by the narrow region, $\bigl||x|- |y| \bigr | \lesssim
1/a<\beta_0$. Due to the latter relation, the distribution of
$|x|- |y|$ can be replaced by the constant,
$1/\sqrt{2\pi}\beta_0$, and the resulting integral can be
calculated. This gives:
\begin{equation}\label{ansdLweak}
\frac{\mathcal{L} (\omega_1)}{\mathcal{L}_0}=
\frac{\sqrt{\pi/2}}{a\beta_0}= \sqrt{\frac\pi{2a^2}}\,
\frac{\omega_1}{\omega_{\text{hf}}}.
\end{equation}
The linear dependence Eq. (\ref{ansdLweak}) of PLDMR amplitude on
$\omega_1$ corresponds to $\propto\sqrt{P_{\text{mw}}}$ dependence
on the microwave power, $P_{\text{mw}}$. This result agrees well
with that of Ref. \onlinecite{RRResonant} and differs from the
earlier predictions of $\propto P_{\text{mw}}$ dependence.
\cite{Eickelkamp, VardenyPRL07}

\subsubsection{Strong resonant drive, $\omega_1\gg
\omega_{\text{hf}}$}

\begin{figure}[t]
\vspace{-0.4cm}
\centerline{\includegraphics[width=95mm,angle=0,clip]{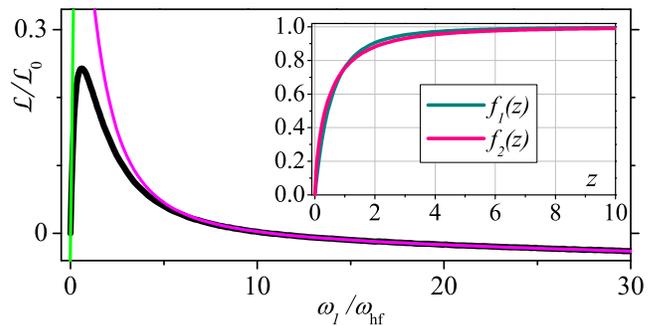}}
\vspace{-0.5cm} \caption{(Color online) The PLDMR amplitude at
zero detuning, $\mathcal{L}(\omega_1)$, found by numerically
solving the Liouville Equation (\ref{sdm}),
is plotted in black, together with the weak-driving asymptote, Eq.
(\ref{ansdLweak}) [green] and the strong-driving asymptote, Eq.
(\ref{ansdLstrong}) [magenta]. The parameters are set to
$w_d\equiv k_d +1/T_{\text{sl}} = 30$ kHz, $k_r=230$ kHz, and
$\omega_{\text{hf}} /2\pi  = 16.8$ MHz, corresponding to the
hyperfine field of 6 Gauss. Inset: Plots of the universal
functions, Eqs. (\ref{f1}), (\ref{f2}). } \label{withasymps}
\end{figure}

In the case of strong drive the second term of Eq.~(\ref{trUans})
is also important.
In this case one typically has $|x|, |y| \ll 1$, and therefore the
approximations,
\begin{equation}\label{strngaprs}
\sin^2(2\phi_{eh})\approx 4y^2,\quad \varepsilon_2 \approx
\omega_1xy,
\end{equation}
can be used in the first and second terms of Eq.~(\ref{trUans}),
respectively. Also, exploiting $2\omega_1 x \gg \sqrt{w_dk_r}$
($\omega_{\text{hf}}\gg \sqrt{w_dk_r}$), in the second term we
replace $\cos^2 (\phi_{eh})$ by 1, neglecting a term $\sim y^2$.
In terms of the constants,
\begin{eqnarray}\label{bK}
&&b=\frac{\sqrt{2w_d(2w_d+k_r)}}{k_r},\quad c=
\frac{\sqrt{w_d(w_d+k_r)}}{\omega_{\text{hf}}}, \nonumber \\
&&\mathcal{B}= \frac{(4w_d+k_r)^2}{k_r^2}, \quad \mathcal{C}=
\frac{4w_d+k_r}{2(w_d+k_r)},
\end{eqnarray}
and the universal functions,
\begin{eqnarray}\label{f1}
&&f_1(z)= z^2\!\!\int\limits_{-\infty}^\infty\!\!
\frac{d\rho}{\sqrt{\pi}} \frac{ e^{-\rho^2}}{\rho^2+z^2} =
\sqrt{\pi} z\exp(z^2)\text{erfc}(z),\qquad \\ \label{f2} &&f_2(z)=
z\!\!\int\limits_0^\infty\!\! d\rho \frac{
e^{-\rho}}{\sqrt{\rho^2+z^2}} = \frac{\pi }2 z \bigl[
H_0(z)-Y_0(z) \bigr],
\end{eqnarray}
where $\text{erfc}(z)$ is the complementary error function, and
$H_0(z)$, $Y_0(z)$ are the zero order Struve and Bessel functions,
respectively, our result reads:
\begin{equation}\label{ansdLstrong}
\frac{\mathcal{L} (\omega_1)}{\mathcal{L}_0} = \mathcal{B}\bigl[1-
f_1\bigl(b\,\omega_1/\omega_{\text{hf}} \bigr)\bigr] - \mathcal{C}
f_2\bigl(c\, \omega_1/\omega_{\text{hf}} \bigr).
\end{equation}
Considering the simple properties of $f_1(z)$ and $f_2(z)$,
plotted in Fig. \ref{withasymps} inset, this equation explains the
decrease of $\mathcal{L}( \omega_1 )$, Fig.~\ref{resdrive}, in
simple terms. First we note that
$b\lesssim 1$, $c\ll 1$, and $\mathcal{B}\sim \mathcal{C}\sim 1$.
Thus, the domain $\omega_1\gtrsim \omega_{\text{hf}}$, next to the
peak of $\mathcal{L}$, is dominated by the first term of
Eq.~(\ref{ansdLstrong}).
The last, Dicke term of Eq.~(\ref{ansdLstrong}) becomes relevant
for $\omega_1\gg \omega_{\text{hf}}$, where the first term
vanishes. Finally, $\mathcal{C}$ gives the saturation value of
$\mathcal{L}( \omega_1 )/\mathcal{L}_0$.

The peak of $\mathcal{L}( \omega_1 )$ occurs between the curves
given by Eqs. (\ref{ansdLweak}) and (\ref{ansdLstrong}). For
$w_d\ll k_r$, entailing small $a\ll 1$, this domain is very narrow
and the position of peak, $\omega_1^{\text{max}}$, is very close
to the intersection of the two curves. From this argument one
finds $\omega_1^{\text{max}}\simeq \omega_{\text{hf}}
a\sqrt{2/\pi}$. The frequency $\omega_1^0$, at which $\mathcal{L}$
becomes 0, can be estimated from the condition that $f_1$ in Eq.
(\ref{ansdLstrong}) is nearly 1. A good estimate for $f_1(z)\simeq
1$ is $z\simeq 5$, corresponding to $\omega_1^0 \simeq 5
\omega_{\text{hf}}/b$. Thus the characteristic values,
$\omega_1^{\text{max}}$ and $\omega_1^0$, are expressed via
$\omega_{\text{hf}}$ and the kinetic parameters, $w_d$ and $k_r$.
This is illustrated in Fig. \ref{withasymps}, where we plot
$\mathcal{L}( \omega_1 )$ found by numerically solving  Eq.
(\ref{sdm}), together with the asymptotes, Eqs. (\ref{ansdLweak}),
(\ref{ansdLstrong}). In Fig.~\ref{withasymps} we used the
parameters inferred for a semiconducting fluorescent polymer,
\cite{DaneNatMat08} implying $w_d\ll k_r$.

\subsection{Lineshape analysis}

\begin{figure}[b]
\vspace{-0.4cm}
\centerline{\includegraphics[width=95mm,angle=0,clip]{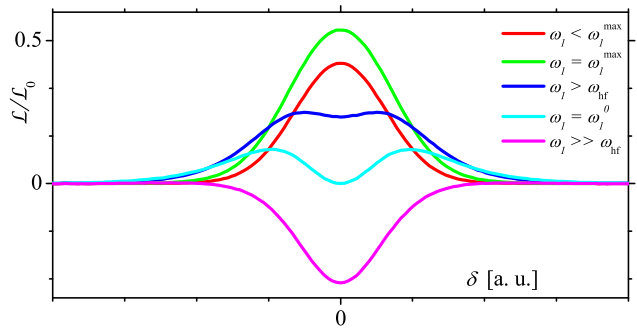}}
\vspace{-0.5cm} \caption{(Color online)  Sketch of lineshapes for
different microwave amplitudes. Upon increasing the drive
amplitude, the line (red) broadens and grows (green), and
subsequently evolves to fully negative (magenta). Emergence of
minimum (blue) manifests the onset of the spin-Dicke effect. Note
that the plots are intended to assist the explanation and do not
reflect any actual dependence.} \label{shapes}
\end{figure}

As illustrated in Fig. \ref{shapes}, the resonance lines can be
divided into four groups by their shapes. At weak drive,
corresponding to the region of initial linear growth in Fig.
\ref{resdrive}, the lineshapes are double Gaussian. In the region
near the maximum in Fig. \ref{shapes}, $\omega_1\lesssim
\omega_{\text{hf}}$, the lines deviate from double Gaussian and
become broader. The next, third group includes the lines with a
minimum at resonance and two mirroring maxima at the sides, is
found for $\omega_1\gtrsim \omega_{\text{hf}}$, and the fourth
type of lines, showing completely negative resonance, appear at
the strongest drives, $\omega_1\gg \omega_{\text{hf}}$. As
discussed shortly, the two latter lineshapes are clear
fingerprints of the spin Dicke effect.

The analytical forms of lineshapes can be found from Eqs.
(\ref{trUans}), (\ref{avtr}), where the local Larmor frequencies
are distributed by Eq. (\ref{GaussDis}), with the non-zero
detuning $\delta$. At weak drive and for the general case of
unequal electron and hole hyperfine coupling strengths, Eq.
(\ref{dLweak}) is valid with a modification of the product,
$\mathcal{P}(x)\mathcal{P}(y)$. The result of the asymptotic
evaluation of the corresponding integral,
\begin{equation}\label{doubleGauss}
\frac{\mathcal{L} (\delta, \omega_1)}{\mathcal{L}_0} =
\frac{\pi\omega_1}{2a} \left (\frac {
e^{-\frac{\delta^2}{2\omega_{\text{hf}, e}^2}}}
{\sqrt{2\pi}\omega_{\text{hf}, e}} + \frac
{e^{-\frac{\delta^2}{2\omega_{\text{hf}, h}^2}}}
{\sqrt{2\pi}\omega_{\text{hf}, h}} \right),
\end{equation}
is the generalization of Eq. (\ref{ansdLweak}) for
$\omega_{\text{hf}, e} \neq \omega_{\text{hf}, h}$. The red curve
in Fig. \ref{shapes} represents such a double Gaussian.
Experimentally, this is the most easily accessible domain of
drive. For stronger microwave strength, while the lineshapes are
still accurately described by Eqs. (\ref{trUans}) and
(\ref{avtr}), only a qualitative analysis will follow.

The green curve in Fig. \ref{shapes} is the line with the largest
amplitude, occurring at $\omega_1= \omega_1^{\text{max}}
<\omega_{\text{hf}}$. It is broader than the double Gaussian Eq.
(\ref{doubleGauss}). Both the red and green curves in Fig.
\ref{shapes} are well described by the first term of Eq.
(\ref{trUans}), meaning that the Dicke subradiant state is not
efficient at this microwave strength.

With the further increase of microwave strength over
$\omega_{\text{hf}}$, the line amplitudes decrease and central
dips appear, as seen for the blue and cyan lines in Fig.
\ref{shapes}. This signifies the onset of the subradiant mode,
whose contribution is negative. The contribution of this mode
overruns the regular term, which in turn becomes progressively
smaller, at yet stronger microwave fields. In Fig. \ref{shapes},
the cyan line shows the situation where the signal is zero exactly
at the resonance, and the magenta line depicts a fully negative
resonance line. The latter represents the form at which the lines
saturate at the strongest drives.

\subsection{Finite spin-lattice relaxation}

From Eqs. (\ref{compHam}) and (\ref{ssrho}) it is easy to see that
the operator $U$ depends on the spin relaxation and the
non-radiative decay only through the combination,
$w_d=k_d+1/T_{\text{sl}}$. At slow spin relaxation the
approximation Eq. (\ref{simpL}) is valid, and therefore, up to an
inessential overall factor, $\mathcal{L}$ also depends on $w_d$,
rather than on $k_d$ or $T_{\text{sl}}$. Thus, in the limit
$1/T_{\text{sl}} \ll k_d$, we encounter the conventional
property of the additive inverse lifetimes.

\begin{figure}[t]
\vspace{-0.4cm}
\centerline{\includegraphics[width=95mm,angle=0,clip]{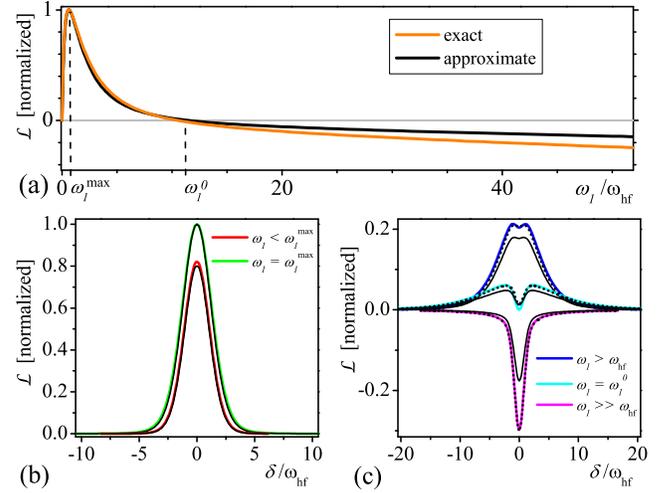}}
\vspace{-0.5cm} \caption{(Color online) Illustration of the role
of spin-lattice relaxation in PPR model. Simulation results are
shown for $T_{\text{sl}}= 40\, \mu$s, $k_d=5$ kHz, and $k_r=230$
kHz (Ref. \onlinecite{DaneNatMat08}). (a) The signal at resonant
driving is plotted from the exact Eq.~(\ref{Lviatr}) with orange,
and from the approximate Eq.~(\ref{simpL}) with black. The plots
are normalized to reach the maximum value of 1. (b) and (c)
Lineshapes for different microwave strengths are plotted in colors
from Eq.~(\ref{Lviatr}), together with the corresponding plots
from the approximate Eq.~(\ref{simpL}) (black lines). The
normalization is the same as in (a). In the domain
$\omega_1\lesssim \omega_1^{\text{max}}$, shown in (b), the
approximate lines are very accurate, reflecting the additive
character of spin relaxation and non-radiative decay.
For $\omega_1> \omega_1^{\text{max}}$, illustrated in (c), the
approximate curves deviate form the exact ones substantially.
Still, approximate lines can be made very close to the exact ones
with individual normalization for each $\omega_1$ (black dots).}
\label{finTsl}
\end{figure}

To scrutinize the regime of intermediate spin relaxation,
$1/T_{\text{sl}} \gtrsim k_d$, we perform numerical simulations
based on the exact formula Eq.~(\ref{Lviatr}).
The results of our numerical analysis show that, besides the
additive feature of inverse lifetimes, the main effect of the spin
relaxation is the overall reduction of the amplitude of
$\mathcal{L}$. However, the latter effect is inessential because
of the overall normalization uncertainty in real experimental
conditions.

Figure \ref{finTsl} compares the finite spin relaxation results
from the hyperfine averaged exact equation (\ref{Lviatr}) with the
outcome of the approximation Eq.~(\ref{simpL}).
The parameters in Fig. \ref{finTsl} are borrowed from Ref.
\onlinecite{DaneNatMat08}, where $1/T_{\text{sl}} \simeq 5 k_d$ is
inferred experimentally. The solid lines in Fig.~\ref{finTsl} are
normalized to ensure the maximal value of 1 for the function,
$\mathcal{L}(\delta, \omega_1)$, which occurs at $\omega_1=
\omega_1^{\text{max}}$ and $\delta=0$, both for the exact and the
approximate solutions.

The plots in Fig.~\ref{finTsl}(a) and (b) clearly indicate very
close results from the exact Eq. (\ref{Lviatr}) and the
approximation Eq.~(\ref{simpL}), thus confirming the additive
character of spin relaxation and non-radiative decay rates for
$\omega_1\lesssim \omega_1^{\text{max}}$ and moderate spin
relaxation, $1/T_{\text{sl}} \gtrsim k_d$. Deviations between the
exact and approximate lines are noticeable in the domain of strong
drive $\omega_1 > \omega_1^{\text{max}}$, Fig.~\ref{finTsl}(c).
Apparently, this could mean that the effect of spin relaxation can
be resolved from that of the non-radiative recombination in the
limit of strong drive. However, the approximate lines can be made
very close to the exact ones upon applying different normalization
factors for different $\omega_1$- values, see
Fig.~\ref{finTsl}(c). Therefore, in order to resolve the spin
relaxation effects,
multiple resonance lines at different strong drive fields are
necessary.

\section{The triplet exciton-polaron quenching (TPQ) model}

Various schemes employing the TPQ mechanism have been invoked in
the literature to date. \cite{ShinarPRL05, Desai07, Koopmans11,
Drew10} Although different in many aspects, all these schemes stem
from the spin dependent reaction between a triplet exciton, TE,
and a polaron, P:
\begin{equation}\label{TPQreact}
\text{TE} + \text{P} \leftrightarrow \text{P}^* + \text{S}_0,
\end{equation}
where S$_0$ stands for a singlet ground state and $*$ denotes a
possibly excited state. While the right hand side of Eq.
(\ref{TPQreact}) is spin doublet, the triplet exciton-polaron
complex (TEP) in the left hand side can form two different spin
multiplets, a quartet and a doublet. Hence the spin dependence of
the reaction (\ref{TPQreact}), which can occur only from the
doublet state of the initial complex. Furthermore, under magnetic
resonance conditions, the TEP spin multiplicity, and therefore the
reaction yield of Eq. (\ref{TPQreact}), can be altered by a
microwave drive.

The reaction (\ref{TPQreact}) does not involve singlet excitons
(SE), and the SE density, $n_{\text{S}}$, becomes sensitive to the
reaction yield because of a quenching of SEs by TEs and polarons.
Ultimately, this quenching facilitates the optical detection of
the microwave-induced reaction yield of Eq. (\ref{TPQreact}). For
simplicity, we will consider the quenching by TEs only, described
by the the rate equation,
\begin{equation}\label{nS}
\partial_t n_{\text{S}} = G_S -R_S n_{\text{S}}
- R_{ST} n_{\text{S}} n_{\text{T}},
\end{equation}
where the SE generation and decay rates, $G_S$ and $R_S$
respectively, are the same as in Eq. (\ref{PPnS}), whereas
$R_{ST}$ is the SE -- TE quenching rate. For the TE density,
$n_{\text{T}}$, one has:
\begin{equation}\label{nT}
\partial_t n_{\text{T}} = G_T -R_T n_{\text{T}}
- R_{ST} n_{\text{T}} n_{\text{S}} -\beta(\delta,
\omega_1)n_{\text{T}},
\end{equation}
where $G_T$ and $R_T$ are respectively the TE generation and decay
rates,
and $\beta(\delta, \omega_1)$ is the rate of the TE population
decline due to the reaction (\ref{TPQreact}), which depends also
on the polaron density, $n_{\text{P}}$.

Under typical conditions, the non-linear terms in Eqs. (\ref{nS}),
(\ref{nT}) are small perturbations, and the  steady-state
densities are quite accurately given by the first two terms in the
rate equation right hand sides:
\begin{equation}\label{tss}
\tilde{n}_{\text{S}} \approx G_S/R_S, \quad \tilde{n}_{\text{T}}
\approx G_T/R_T.
\end{equation}
Note that the description Eqs. (\ref{nS})-(\ref{tss}) is valid for
not very strong photoexcitation power $P_L$, ensuring a linear
regime with $\tilde{n}_{\text{S}}\propto P_L$  ($G_S \propto
P_L$).

In order to describe the magnetic field effects, higher order
corrections to Eq. (\ref{tss}) must be considered. From Eqs.
(\ref{nS}), (\ref{nT}) we find:
\begin{eqnarray}\label{SSnS}
\tilde{n}_{\text{S}} = \sqrt{\left( \frac{R_T+\beta}{2R_{ST}} +
\frac{G_T -G_S}{2R_S} \right)^2 + \frac
{G_S(R_T+\beta)}{R_SR_{ST}} }&&
\nonumber \\
-\left(\frac{R_T+\beta}{2R_{ST}} + \frac{G_T -G_S}{2R_S} \right).
&&
\end{eqnarray}
The microwave-induced change of population, $\Delta
\tilde{n}_{\text{S}} = \tilde{n}_{\text{S}}(\omega_1) -
\tilde{n}_{\text{S}}(0)$, is found from Eq. (\ref{SSnS}) to be
\begin{equation}\label{dltn}
\Delta\tilde{n}_{\text{S}} = \frac{\tilde{n}_{\text{S}}
\tilde{n}_{\text{T}} R_{ST}} {R_SR_T}\bigl[
\beta(\delta,\omega_1)- \beta(0) \bigr],
\end{equation}
where we have used the leading order results, Eq. (\ref{tss}), and
the relation, $G_S\gg G_T$.

We derive $\beta (\delta, \omega_1)$ in Appendix \ref{AppTPQ} from
the stochastic Liouville approach, by assuming that the steady
state TEP generation rate is given by the product, $\lambda\,
\tilde{n}_{\text{P}}\tilde{n}_{\text{T}}$, where $\lambda$ is a
constant determined by the TEP formation cross section. We get:
\begin{equation}\label{formalp}
\beta (\delta, \omega_1) = \lambda\, \tilde{n}_{\text{P}} \Gamma(
\delta, \omega_1 ),
\end{equation}
where $\Gamma(\delta, \omega_1 )$ is governed by the TEP spin
dynamics and recombination. Thus, the (normalized) optically
detected signal, $\Delta\mathcal{I}/\mathcal{I} =
\Delta\tilde{n}_{\text{S}}/\tilde{n}_{\text{S}}$, is given by
\begin{equation}\label{nrmdltn}
\frac{\Delta\mathcal{I}(\delta,\omega_1)}{\mathcal{I}} =
\tilde{n}_{\text{P}} \tilde{n}_{\text{T}} \frac{\lambda \, R_{ST}}
{R_SR_T} \bigl[ \,\Gamma(\delta,\omega_1) - \Gamma(0) \bigr].
\end{equation}

\begin{figure}[t]
\centerline{\includegraphics[width=60mm,angle=0,clip]{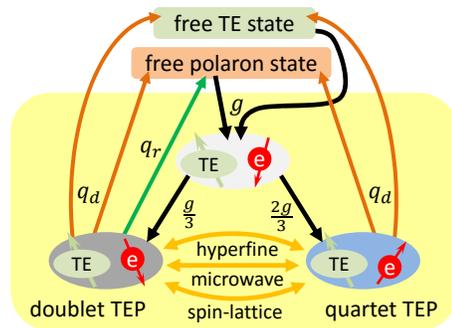}}
\caption{(Color online) Schematics of the processes involved in
the TPQ model. The specifics of TPQ is that the microwave drive,
together with the hyperfine coupling and spin-lattice relaxation,
couples the TEP spin levels. The color code coincides with that in
Fig.~\ref{PPRschem}. The arrangement of the states is of no
importance.} \label{TPQschem}
\end{figure}

For the following discussion we present the zero-detuning result
for $\Gamma( \omega_1 )\equiv \Gamma(0,\omega_1)$, established in
the limit of weak dissociation and recombination, and negligible
coupling between the polaron and TE spins (see Appendix
\ref{AppTPQ}). Figure \ref{TPQschem} depicts the processes
underlying the TPQ model. It includes the TEP generation rate,
$g$, the dissociation rate, $q_d$, the rate of the reaction
Eq.~(\ref{TPQreact}) from the doublet manifold, $q_r$. Not shown
in Fig. \ref{TPQschem} are the TEP spin-lattice relaxation time,
$T_{\text{sl}}$, and the polaron hyperfine coupling magnitude,
$\omega_{\text{hf}}$. In the limit of long spin coherence times,
$T_{\text{sl}}> 1/q_d$,  and slow dissociation and recombination,
$q_d$, $q_r \ll \omega_{\text{hf}}$, we find $\Gamma( \omega_1 ) -
\Gamma(0)= \Gamma_0 f_1(\omega_1/\omega_s)$, and therefore
\begin{equation}\label{TPQdltn}
\frac{\Delta\mathcal{I}(\omega_1)}{\mathcal{I}} =
\tilde{n}_{\text{P}} \tilde{n}_{\text{T}} \frac{\lambda \, R_{ST}
\Gamma_0} {R_SR_T} f_1(\omega_1/\omega_s),
\end{equation}
where
\begin{eqnarray}\label{Gom}
&&\Gamma_0= \frac{2\,q_d\,q_r^2}{3v_d(3v_d+q_r)(3v_d+2q_r)},
 \\
&&\omega_s = \omega_{\text{hf}} \frac{\sqrt{6v_d(3v_d+2q_r)}}
{3v_d+q_r}, \qquad v_d= q_d+1/T_{\text{sl}},\nonumber
\end{eqnarray}
are constants. The function $f_1(z)$ appeared earlier in PPR
model, see Eq. (\ref{f1}). It is plotted in Fig.~\ref{withasymps}
inset. It grows from zero linearly, and
saturates to unity at $z>1$. This translates into the initially
linear growth of $\Delta\mathcal{I}/\mathcal{I}$, and saturation
to $\tilde{n}_{\text{P}} \tilde{n}_{\text{T}} \Gamma_0
\bigl(\lambda\, R_{ST}/R_SR_T \bigr)$ at $\omega_1> \omega_s$.

Note that Eq. (\ref{TPQdltn}) represents the contribution from
only one species of polarons.
To account for the other, charge-conjugated species, a term
similar to that in the right hand side of Eq. (\ref{TPQdltn}) must
be included, with the corresponding values of $\lambda$,
$\Gamma_0$, $R_{ST}$, $R_S$, $R_T$, and $\omega_s$.

\section{Discussion and summary}

The present study of the magnetic resonance-induced variation of
singlet exciton recombination is based on the description of spin
dynamics and recombination by means of stochastic Liouville
equations. For the PPR model, we have demonstrated a solution
method yielding the answer in terms of the eigenvalues of $4\times
4$ complex Hamiltonian, instead of the solution of effectively $10
\times 10$ complex linear system of stochastic Liouville
equations. Analytical results supported by the direct numerical
solution of stochastic Liouville equations are found in the limit
of weak singlet recombination. The microwave-induced spin Dicke
effect, stemming from the back-action of recombination on the
quantum dynamics of spin pairs, is identified and described
quantitatively.

We have considered a spin-lattice relaxation, uniform with respect
to the spin multiplicity. If the relaxation time is not too short,
the main effect of this relaxation is additive to that of the
dissociation and non-radiative recombination of the polaron pairs.
We have shown that it can influence the resonance lines only at
strong drive, whereas at weak drive it leads to the overall
scaling of resonance amplitudes. Note in passing that our approach
naturally takes into account the dominant $T_2$- processes,
originating from the random hyperfine interaction.

Our analysis excludes the exchange and dipolar interactions
between the spin pairs, although these interactions can be readily
included in the presented perturbative scheme. This is done for
the sake of simplicity, since our direct numerical simulations
show that the effect of these interactions is minor, given that
they do not exceed the average hyperfine coupling strength.
\cite{Kipp15}

The TPQ model is treated along the same lines. However,
calculations in this case are greatly simplified due to the
presence of relatively strong zero-field splitting of triplets,
making these states off resonance.

Concurring results are found from the PPR and TPQ models at weak
drive. Namely, if the TPQ reaction Eq.~(\ref{TPQreact}) is
equally probable for the electron and hole polarons, the
lineshapes from the two models are the same for $\omega_1\ll
\omega_{\text{hf}}$, and are given by the sum of two Gaussians,
Eq. (\ref{doubleGauss}).

More importantly, we uncover two substantial differences in the
predictions of the PPR and the TPQ models. First and foremost, the
dissimilar dependence of the microwave-induced signal on the
steady-state densities,
\begin{equation}\label{twDM}
\Delta \mathcal{I}\propto \tilde{n}_{\text{P}}^2 \quad
\text{(PPR)}, \qquad \Delta \mathcal{I}\propto
\tilde{n}_{\text{S}} \tilde{n}_{\text{P}}\tilde{n}_{\text{T}}
\quad \text{(TPQ)},
\end{equation}
cf. Eqs. (\ref{NormdL}) and (\ref{TPQdltn}), respectively, leads
to the remarkably different results for the dependence of $\Delta
\mathcal{I}/\mathcal{I}$ on the photoexcitation power, $P_L$. Far
from saturation at high $P_L$ it is reasonable to expect that
$\tilde{n}_{\text{S}} \propto P_L$, $\tilde{n}_{\text{P}} \propto
P_L$, and $\tilde{n}_{\text{T}} \propto P_L$ crossing over to
$\tilde{n}_{\text{T}}\propto P_L^2$ (the position of crossover
depends on the efficiency of the intersystem crossing from SE to
TE and the TE generation from non-geminate polaron pairs
\cite{List01}). For the TPQ model, this results in
$\Delta \mathcal{I}/\mathcal{I} \propto P_L^2$ to $P_L^3$, in
contrast to the PPR prediction, $\Delta \mathcal{I}/ \mathcal{I}
\propto P_L$.

The second important difference between the predictions of the two
models comes from the fact that, at the polaron spin-$1/2$
resonance, TEs are mainly off resonance because of a relatively
strong zero-field splitting. As a result, the lineshapes and the
saturation behavior from the two models are different at strong
drive. Specifically, the TPQ leads to the resonance lines
featuring a single maximum, and relatively fast saturation of
$\Delta \mathcal{I}/\mathcal{I}$ to positive values at
$\omega_1\sim \omega_{\text{hf}}$, much like in the ordinary ESR.
In contrast, the PPR model predicts resonance lines with two
maxima around the central dip at $\omega_1\gtrsim
\omega_{\text{hf}}$, evolving into the completely negative
resonance at $\omega_1\gg \omega_{\text{hf}}$, where $\Delta
\mathcal{I}/\mathcal{I}$ saturates to negative values (see Figs.
\ref{resdrive} and \ref{shapes}). These differences are of
relevance for resolving the contributions of the two mechanisms
experimentally, via continuous wave PLDMR measurements.

Finally, we note that this study did not address the predictions
for the observables which are measured using the double modulation
(DM) PLDMR technique. \cite{ShinarPRL05, ShinarPRB05} In this
technique the laser excitation power is modulated at certain
frequency $f_L$, and a lock-in amplifier filters out the delayed
photoluminescence that is slower than $f_L$. Therefore, by the
design, the DM-PLDMR measures only the prompt component of the
photoluminescence. The results obtained using the DM-PLDMR
\cite{ShinarPRL05, ShinarPRB05} are independent of $f_L$ up to
$100$ kHz. In this regard, we would like to note that our Eq.
(\ref{twDM}) offers a certain prediction for DM-PLDMR. Namely, the
proportionality of $\Delta \mathcal{I}$ to $\tilde{n}_{\text{S}}$
renders the interpretation of the DM-PLDMR results
\cite{ShinarLPR12, ShinarPRL05, ShinarPRB05} in favor of the TPQ
model. Further theoretical studies aimed at more quantitative
predictions for DM-PLDMR are underway.

\section*{Acknowledgments}

We thank H. Malissa for helpful discussions. Work at the Ames
Laboratory was supported by the US Department of Energy, Office of
Science, Basic Energy Sciences, Division of Materials Sciences and
Engineering. The Ames Laboratory is operated for the US Department
of Energy by Iowa State University under Contract No.
DE-AC02-07CH11358. M. R. acknowledges the support of the US
Department of Energy, Office of Basic Energy Sciences, Grant No.
DE-FG02-06ER46313.

\appendix

\section{}

\label{AppPP}

In this Appendix we investigate the steady-state
Liouville equation for the PPR model, Eq. (\ref{sdm}).

Because of the non-Hermitian character of the Hamiltonian
$\mathcal{H}$, Eq. (\ref{compHam}), the calculation of
$\text{tr}\, U$ is specific. We introduce the eigenvectors and
eigenvalues, $\mathcal{H} |\psi_\alpha\rangle = \varepsilon_\alpha
|\psi_\alpha\rangle$, $\alpha=1,..,4$. As the non-Hermitian
Hamiltonian $\mathcal{H}$ is symmetric, the conjugated equation $
\langle\psi_\alpha| \mathcal{H} = \varepsilon_\alpha
\langle\psi_\alpha|$ holds for $\langle\psi_\alpha|=|\psi_\alpha
\rangle^\intercal$, where the superscript $\intercal$ means the
transpose without a complex conjugation.
We normalize the eigenvectors with respect to this conjugation, so
that $\langle\psi_\alpha| |\psi_\alpha\rangle =1$. It is also easy
to check that the eigenvectors are orthogonal;
$\langle\psi_\alpha| |\psi_\beta\rangle\equiv \sum_{n=1}^4
\psi_\alpha(n) \psi_\beta(n) =0$, if $\varepsilon_\alpha\neq
\varepsilon_\beta$, whereas the degenerate case can be handled in
the standard way, by choosing orthogonal vectors in the degenerate
subspace. Thus $\{|\psi_\alpha\rangle\}_{\alpha=1}^4$ can be made
a complete orthonormal set. This ensures the partition of unity,
$\sum_{\alpha=1}^4|\psi_\alpha\rangle\langle \psi_\alpha| =
\mathbb{1}$, yielding $\text{tr}\, U = \sum_{\alpha=1}^4\langle
\psi_\alpha|U|\psi_\alpha\rangle $. The complex conjugate vectors,
$|\psi_\alpha^*\rangle\equiv |\psi_\alpha \rangle^*$, obeying
$\mathcal{H}^* |\psi_\alpha^*\rangle = \varepsilon _\alpha^*
|\psi_\alpha^*\rangle$, form another orthonormal set, in general
different from $\{|\psi_\alpha\rangle\}_{\alpha=1}^4$. With these
conventions, from Eq. (\ref{ssrho}) we write:
\begin{equation}\label{trU}
\text{tr}\, U = \sum_{\alpha,\beta=1}^4 \frac{\langle\psi_\alpha|
|\psi_\beta^*\rangle\langle\psi_\beta^*| |\psi_\alpha\rangle}{i(
\varepsilon_\alpha - \varepsilon_\beta^*)}.
\end{equation}
Treating the recombination term of the Hamiltonian
(\ref{compHam}), $V=-i(k_r/2)\Pi_S$, as a perturbation, we get:
\begin{equation}\label{numer}
\langle\psi_\alpha| |\psi_\beta^*\rangle\langle\psi_\beta^*|
|\psi_\alpha\rangle = \delta_{\alpha \beta}+ \mathcal{O}\bigl
(k_r/\Omega_{e,h} \bigr)^2.
\end{equation}
and
\begin{equation}\label{denom}
\varepsilon_\alpha = \epsilon_\alpha -i(w_d/2) +V_{\alpha \alpha}
+ \mathcal{O}\bigl (k_r^2/\Omega_{e,h} \bigr),
\end{equation}
where $\epsilon_\alpha$ are the eigenvalues of the Hamiltonian
$H$, Eq.~(\ref{Hamil}). By observing that the $\sim
k_r^2/\Omega_{e,h}$ correction in $\varepsilon_\alpha$ is real,
from Eqs. (\ref{trU}) - (\ref{denom}) we infer that omitting the
inexplicit terms in Eqs. (\ref{numer}), (\ref{denom}) induces only
$\sim (k_r/\Omega_{e,h} )^2 \text{tr}\, U$ error in $\text{tr}\,
U$. Therefore, rather accurate results can be found by completely
neglecting the eigenvector corrections and keeping only the
leading corrections to $\varepsilon_\alpha$. This is as much as we
get from Eq. (\ref{trU}), because the degeneracy of $H$ makes the
simple perturbation calculation inefficient.

Still, the explicit form of the unperturbed eigenstates of
$\mathcal{H}$, $|\varphi_\alpha \rangle$, which are the
eigenvectors of $H$, is needed. In the absence of the exchange and
dipolar interactions, the
individual electron ($\mu=e$) and hole ($\mu=h$) polaron
eigenstates are:
\begin{eqnarray}\label{indsts}
&&|\Uparrow\rangle_\mu =\cos\phi_\mu |\uparrow\rangle_\mu +
\sin\phi_\mu |\downarrow\rangle_\mu, \nonumber\\
&&|\Downarrow\rangle_\mu = \sin\phi_\mu |\uparrow\rangle_\mu -
\cos\phi_\mu |\downarrow\rangle_\mu,
\end{eqnarray}
where $|\uparrow\rangle_\mu$, $|\downarrow\rangle_\mu$ are the
electron and hole polaron spin up and down states with the
quantization axes along $\hat{\mathbf{z}}$,  and $\tan
2\phi_\mu=\omega_1/\omega_\mu$
is introduced. Then we have:
\begin{eqnarray}\label{eigs}
&&|\varphi_1 \rangle = |\Uparrow\rangle_e |\Uparrow\rangle_h,
\quad |\varphi_2 \rangle = |\Uparrow\rangle_e
|\Downarrow\rangle_h,\nonumber\\
&&|\varphi_{3} \rangle = |\Downarrow\rangle_e|\Uparrow\rangle_h,
\quad |\varphi_4 \rangle = |\Downarrow\rangle_e
|\Downarrow\rangle_h.
\end{eqnarray}
The matrix $V_{\alpha \beta}= \langle \varphi_\alpha | V|
\varphi_\beta\rangle$, Eq. (\ref{Vab}) in the main text, is found
by drawing in the singlet-triplet base states,
\begin{eqnarray}\label{ST}
&&|T_+\rangle = |\uparrow\rangle_e |\uparrow\rangle_h, \quad
|T_-\rangle = |\downarrow\rangle_e |\downarrow\rangle_h, \nonumber\\
&&|T_0\rangle =
\frac1{\sqrt{2}}\bigl(|\uparrow\rangle_e|\downarrow\rangle_h +
|\downarrow\rangle_e|\uparrow\rangle_h \bigr), \nonumber\\
&&|S\rangle = \frac1{\sqrt{2}}
\bigl(|\uparrow\rangle_e|\downarrow\rangle_h -
|\downarrow\rangle_e|\uparrow\rangle_h \bigr),
\end{eqnarray}
and calculating the scalar products of $|\varphi_\alpha \rangle$
with $|S\rangle$.

The degeneracy of $H$, controlling the strong drive regime,
corresponds to $\phi_\mu\approx \pi/4$. Then Eqs.~(\ref{indsts}) -
(\ref{ST}) give:
\begin{eqnarray}\label{phiST}
&&|\varphi_1 \rangle \approx \frac12\bigl( |T_+ \rangle
+|T_-\rangle + \sqrt{2}\, |T_0\rangle \bigr ) \nonumber\\
&&|\varphi_2 \rangle \approx \frac12\bigl( |T_+ \rangle
+|T_-\rangle - \sqrt{2}\, |S\rangle \bigr ) \nonumber\\
&&|\varphi_3 \rangle \approx \frac12\bigl( |T_+ \rangle
+|T_-\rangle + \sqrt{2}\, |S\rangle \bigr ) \nonumber\\
&&|\varphi_4 \rangle \approx \frac12\bigl( |T_+ \rangle
+|T_-\rangle - \sqrt{2}\, |T_0\rangle \bigr ),
\end{eqnarray}
The vectors $|\varphi_1 \rangle$ and $|\varphi_4 \rangle$ are
always the eigenstates of $\mathcal{H}$ to the leading order,
whereas $|\varphi_2 \rangle$ and $|\varphi_3 \rangle$ are not such
in the vicinity of the degeneracy of $H$. On the other hand, the
eigenvectors of $\tilde{\mathcal{H}}=H -i(w_d/2)\mathbb{1} +
\tilde{V}$, where $\tilde{V}$ is given by Eq. (\ref{tildeV}) in
the main text, are always the leading order eigenstates of
$\mathcal{H}$. In addition, $\mathcal{H}$ and
$\tilde{\mathcal{H}}$ have the same eigenvalues within the first
subleading order. Thus, by virtue of the arguments presented after
Eq. (\ref{denom}), one can replace $\mathcal{H}$ by
$\tilde{\mathcal{H}}$ in Eq. (\ref{ssrho}) of the main text and
calculate $\text{tr}\, U$ with a satisfactory precision,
regardless of the degeneracy. The calculation of $\text{tr}\, U$
is facilitated by fact that $\tilde{\mathcal{H}}$ is diagonal in
the $(\varphi_1, \varphi_4)$ subspace, whereas the contribution of
the $(\varphi_2, \varphi_3)$ subspace can be found by using the
formula,
\begin{equation}\label{expHH}
e^{i(x\sigma_x +y\sigma_z)}= \cos r + i(x \sigma_x + y \sigma_z)
\sin r /r,
\end{equation}
where $\sigma_{x,z}$ are the Pauli matrices, and $r = \sqrt{x^2+
y^2}$. This gives:
\begin{eqnarray}\label{trUans1}
&&\text{tr}\, U = \frac4{k_r\sin^2(\phi_{eh}) +2w_d}+
\frac4{k_r\cos^2(\phi_{eh}) +2w_d} \\
&&+\frac{k_r^2\cos^4(\phi_{eh}) } {\bigl[k_r\cos^2(\phi_{eh})
+2w_d\bigr]\bigl[4 \varepsilon_2^2 +w_d(k_r\cos^2(\phi_{eh})
+w_d)\bigr]},  \nonumber
\end{eqnarray}
where the first term comes from the $(\varphi_1, \varphi_4)$, and
the last two terms -- from the $(\varphi_2, \varphi_3)$ manifolds.
Combining the first two terms gives Eq. (\ref{trUans}) of the main
text.

\section{}

\label{AppTPQ}

In this Appendix we derive the rate $\beta(\delta, \omega_1)$,
introduced in Eq. (\ref{nT}) of the main text, in the limit of
negligible exchange and dipolar coupling between the TE and
polaron spins, and weak dissociation and recombination.

The basis spin states of a TEP complex can be given through the
direct product of a triplet and doublet states, as well as through
the direct sum of a quartet and doublet multiplets, via the
Clebsh-Gordan coefficients. In terms of the components, $T_0$,
$T_{\pm}$, representing triplet exciton with spin projection $0$
and $\pm 1$, respectively, and $\uparrow$, $\downarrow$ for
polaron spin $\pm 1/2$,
\begin{eqnarray}\label{quartanddoub}
&&|T_+ \uparrow \rangle =Q_{3/2}, \nonumber\\
&&|T_+ \downarrow \rangle =\sqrt{1/3}\, Q_{1/2} + \sqrt{2/3}\, D_{1/2}, \nonumber\\
&&|T_0 \uparrow \rangle =\sqrt{2/3}\, Q_{1/2} - \sqrt{1/3}\, D_{1/2}, \nonumber\\
&&|T_0 \downarrow \rangle =\sqrt{2/3}\, Q_{-1/2} + \sqrt{1/3}\, D_{-1/2}, \nonumber\\
&&|T_- \uparrow \rangle =\sqrt{1/3}\, Q_{-1/2} - \sqrt{2/3}\, D_{-1/2}, \nonumber\\
&&|T_- \downarrow \rangle =Q_{-3/2},
\end{eqnarray}
where $Q_k$ and $D_k$ are the quartet and doublet states with the
spin projection $k$.

The $6\times 6$ spin density matrix of an ensemble of TEP
complexes,  $\varrho$, can be treated by a stochastic Liouville
equation. Formally, the Liouville equation for $\varrho$ is found
by rewriting Eq. (\ref{sLe}) with the following modifications:

({\it i}) The rotating-frame spin Hamiltonian $H$ is given by
\begin{equation}\label{TPQham}
H = \omega_{\text{P}} S_z + H_{0,T} + \omega_1(I_x +S_x),
\end{equation}
where $\omega_{\text{P}}$ is the polaron Larmor frequency,
$\mathbf{S}=1/2$ and $\mathbf{I}=1$ are the polaron and TE spin
operators, and
\begin{equation}\label{H0trp}
H_{0,T} = \omega_T I_z + \mathcal{D}\bigl(I_z^2-I(I+1)/3\bigr),
\end{equation}
is the free TE spin Hamiltonian with the TE Larmor frequency,
$\omega_T=\gamma_T \hbar(B_0+ b_T^z) -\omega$, and the axial
zero-field splitting
parameter, $\mathcal{D}$ (the transverse  zero-field splitting is
neglected in the secular approximation).
We take equal gyromagnetic ratios \cite{ShinarLPR12} of TEs and
polarons; $\gamma_T=\gamma_{e,h}$.

({\it ii}) The generation rate, second term in Eq. (\ref{sLe}), is
replaced by $(g/6)\mathbb{1}$, implying equal probability of the
TEP generation in 6 different spin states. Similarly, the factors
of $1/4$ in the last terms of Eqs. (\ref{Rsl}) and (\ref{rewsLe})
are replaced by $1/6$.

({\it iii}) The dissociation - recombination rates
are denoted respectively by $q_d$ and $q_r$, so that Eq.
(\ref{Rgr}) goes into
\begin{equation}\label{TPQdr}
\mathcal{R}_{\text{dr}}\{\varrho\}_{\alpha \beta}= -
q_d\varrho_{\alpha \beta} -\frac{q_r}2\! \sum_{\sigma = \pm
\frac12}\! (\delta_{\alpha D_\sigma} + \delta_{D_\sigma
\beta})\varrho_{\alpha \beta},
\end{equation}
implying a recombination from the doublet TEP states.

({\it iv}) The second term in Eq. (\ref{compHam}) is written as
$-i(v_d/2)\mathbb{1}$, where $v_d=q_d+1/T_{\text{sl}}$, and, more
importantly, the operator $\Pi_S$ is replaced by the projection
operator onto the doublet,
\begin{equation}\label{PD}
\Pi_D = | D_{1/2}\rangle\langle D_{1/2}| + |
D_{-1/2}\rangle\langle D_{-1/2}|.
\end{equation}

We further assume that the TEP generation rate is proportional to
$n_{\text{T}}$ and $n_{\text{P}}$; $g = \lambda n_{\text{T}}
n_{\text{P}}$, and that after dissociation of a TEP complex, the
constituent TE returns into the state described by $n_{\text{T}}$
(see Fig. \ref{TPQschem}). The latter assumption allows us to
write the TEP counterpart of Eq. (\ref{tr}): $\beta
\tilde{n}_{\text{T}} = g - q_d\, \text{tr}\tilde{\varrho}$, where
$g$ gives the rate of the decrease of $n_{\text{T}}$ due to the
generation of TEP,
and the last term reflects the increase of $n_{\text{T}}$ because
of dissociation. By virtue of the full analogy with PPR model, see
Eq.~(\ref{Lviatr}), we write:
\begin{equation}\label{alpviatr}
\beta (\delta, \omega_1) = \lambda
\tilde{n}_{\text{P}}\Gamma(\delta, \omega_1),\quad \Gamma=
\left\langle \frac{1 - \frac{v_d}6\text{tr}\bar{U}} { 1 -
\frac1{6T_{\text{sl}}}\text{tr} \bar{U}}\right\rangle
_{\!\text{hf}},
\end{equation}
where $\bar{U}$ is given by the TPQ counterpart of Eq.
(\ref{ssrho}).

Despite the TE and polaron gyromagnetic ratios are taken to be the
same, \cite{ShinarLPR12} the majority of TE spins are off
resonance because of the relatively strong zero-field splitting.
In Eq. (\ref{TPQham}) we have
$\mathcal{D}=\mathcal{D}_0(3\cos^2\theta-1)/2$, where $\theta$ is
the angle between the zero-field tensor principal $z$- axis and
$\hat{\mathbf{z}}$, and $\mathcal{D}_0\gtrsim 500$ G is measured
for several polymer PPV derivatives. \cite{ShinarLPR12} The
portion of near-resonance TEs is $\sim\omega_1/\mathcal{D}_0$, and
most of these TEs are still off resonance because of the non-zero
TE hyperfine coupling. Therefore, we calculate $\text{tr}\bar{U}$
to the leading order in $\omega_1/\mathcal{D}_0$ and
$\omega_{\text{hf}}/\mathcal{D}_0$, corresponding to the
perturbation,
\begin{equation}\label{barV}
\bar{V}=-i(q_r/2)\Pi_D +\omega_1(I_x +S_x)
\end{equation}
The left-hand side states
in Eq. (\ref{quartanddoub}) are then the unperturbed eigenstates.
Because of the $\sim\mathcal{D}_0$ energy splitting between $|T_+
\uparrow, \downarrow\rangle$ and $|T_0 \uparrow,
\downarrow\rangle$, and  between $|T_- \uparrow,
\downarrow\rangle$ and $|T_0 \uparrow, \downarrow\rangle$, the
matrix elements of $\bar{V}$, relevant to the leading order, are
those between the same $T_i$-states, explicitly given by
\begin{equation}\label{barV}
\bar{V} \simeq \left(\!\begin{array}{cccccc}
0&\frac{\omega_1}2&\cdot&\cdot&\cdot&\cdot\\
\frac{\omega_1}2 &-\frac{iq_r}3&\cdot&\cdot&\cdot&\cdot\\
\cdot&\cdot&-\frac{iq_r}6&\frac{\omega_1}2&\cdot&\cdot\\
\cdot&\cdot&\frac{\omega_1}2&-\frac{iq_r}6&\cdot&\cdot\\
\cdot&\cdot&\cdot&\cdot&-\frac{iq_r}3&\frac{\omega_1}2\\
\cdot&\cdot&\cdot&\cdot&\frac{\omega_1}2&0
\end{array}\!\right)\! .
\end{equation}
The structure of the matrix (\ref{barV}) indicates that the system
reduces to three two-level subsystems, which are decoupled in the
leading order. Further calculation of $\text{tr}\bar{U}$ is done
by using Eq. (\ref{expHH}) for each of the three subsystems,
yielding
\begin{equation}\label{trUbar}
\text{tr}\bar{U}= \frac2{v_{dr}}+ \frac{2v_{dr} \bigl[v_{dr}^2+
\omega_{\text{P}}^2+ \omega_1^2\bigr]} {\bigl[v_{dr}^2+
\omega_{\text{P}}^2\bigr]\bigl[v_{dr}^2-q_r^2/9 \bigr] +v_{dr}^2
\omega_1^2},
\end{equation}
where $v_{dr}= v_d+q_r/3$. The hyperfine average in
Eq.~(\ref{alpviatr}) is over the Gaussian distribution of
$\omega_{\text{P}}$, given by Eq.~(\ref{GaussDis}).
By expanding the denominator in Eq.~(\ref{alpviatr}) over the
small $\text{tr}\bar{U}/T_{\text{sl}}$ and using Eq.
(\ref{trUbar}) we get
\begin{equation}\label{fom1}
\Gamma( \omega_1 ) - \Gamma(0)= \Gamma_0\!
\left(\frac{\omega_1}{\omega_s} \right)^2\!
\int\limits_{-\infty}^\infty \frac{dz}{\sqrt{\pi}}
\frac{\exp(-z^2)}{z^2 + (\omega_1/\omega_s)^2},
\end{equation}
found by neglecting $v_{dr}^2/2\omega_{\text{hf}}^2\ll1$ in the
denominator. This integral happens to coincide with Eq. (\ref{f1})
for $f_1(z)$, leading to the result, Eq. (\ref{TPQdltn}).


\begin{references}


\bibitem{Cavenett81} B. C. Cavenett, ``Optically detected magnetic resonance (O.D.M.R.)
investigations of recombination processes in semiconductors'',
Adv. Phys. {\bf 30}, 475 (1981).

\bibitem{Street82} R. A. Street, ``Recombination in $\alpha$-Si: H: Spin-dependent
effects'', Phys. Rev. B {\bf 26}, 3588 (1982).

\bibitem{Depinna82} S. Depinna, B. C. Cavenett, I. G. Austin, T. M. Searle, M. J.
Thompson, J. Allison, and P. G. L. Comberd, ``Characterization of
radiative recombination in amorphous silicon by optically detected
magnetic resonance: Part I'', Philos. Mag. B {\bf 46}, 473 (1982).

\bibitem{ShinarLPR12} J. Shinar, ``Optically detected magnetic resonance studies of
luminescence-quenching processes in $\pi$-conjugated materials and
organic light-emitting devices'', Laser Photonics Rev. {\bf 6},
767 (2012).

\bibitem{Wohlgenannt01} M. Wohlgenannt, K. Tandon, S. Mazumdar, S. Ramasesha,
and Z. V. Vardeny, ``Formation cross-sections of singlet and
triplet excitons in $\pi$-conjugated polymers'', Nature (London)
{\bf 409}, 494 (2001).

\bibitem{ShinarPRL05} M.-K. Lee, M. Segal, Z. G. Soos, J. Shinar,
and M. A. Baldo, ``Yield of singlet excitons in organic
light-emitting devices: A double modulation
photoluminescence-detected magnetic resonance study'', Phys. Rev.
Lett. {\bf 94}, 137403 (2005).

\bibitem{VardenyPRL07} C. G. Yang, E. Ehrenfreund, and Z. V.
Vardeny,``Polaron spin-lattice relaxation time in $\pi$-conjugated
polymers from optically detected magnetic resonance'', Phys. Rev.
Lett. {\bf 99}, 157401 (2007).


\bibitem{ShinarPRB05} M. Segal, M. A. Baldo, M. K. Lee, J. Shinar,
and Z. G. Soos, ``Frequency response and origin of the spin-$1/2$
photoluminescence-detected magnetic resonance in a
$\pi$-conjugated polymer'', Phys. Rev. B {\bf 71}, 245201 (2005).

\bibitem{ShinarPRB15} Y. Chen, M. Cai, E. Hellerich, R. Shinar,
and J. Shinar, ``Origin of the positive spin-$1/2$
photoluminescence-detected magnetic resonance in $\pi$-conjugated
materials and devices'', Phys. Rev. B {\bf 92}, 115203 (2015).

\bibitem{VardenyPRB08} C. G. Yang, E. Ehrenfreund, F. Wang, T. Drori,
and Z. V. Vardeny, ``Spin-dependent kinetics of polaron pairs in
organic light-emitting diodes studied by electroluminescence
detected magnetic resonance dynamics'', Phys. Rev. B {\bf 78},
205312 (2008).

\bibitem{BoehmeJacs11} S.-Y. Lee, S.-Y. Paik, D. R. McCamey, J. Yu,
P. L. Burn, J. M. Lupton, and C. Boehme, ``Tuning hyperfine fields
in conjugated polymers for coherent organic spintronics'', J. Am.
Chem. Soc. {\bf 133}, 2019 (2011).

\bibitem{RRSlow} R. C. Roundy and M. E. Raikh, ``Slow dynamics of
spin pairs in a random hyperfine field: Role of inequivalence of
electrons and holes in organic magnetoresistance'', Phys. Rev. B
{\bf 87}, 195206 (2013).

\bibitem{RRResonant} R. C. Roundy and M. E. Raikh,
``Organic magnetoresistance under resonant ac drive'', Phys. Rev. B {\bf 88},
125206 (2013).

\bibitem{Lvov82} V. S. L'vov, L. S. Mima, and O. V. Tretyak, ``Investigation of
spin-dependent recombination in semiconductors'', Sov. Phys.
JETP {\bf 56}, 897 (1982).

\bibitem{Barabanov96} A. V. Barabanov, O. V. Tretiak, and V. A.
L'vov,``Complete theoretical analysis of the Kaplan-Solomon-Mott
mechanism of spin-dependent recombination in semiconductors'',
Phys. Rev. B {\bf 54}, 2571 (1996).

\bibitem{Waters15} D. P. Waters, G. Joshi, M. Kavand, M. E. Limes, H. Malissa, P. L. Burn,
J. M. Lupton, and C. Boehme, ''The spin-Dicke effect in OLED
magnetoresistance'', Nat Phys {\bf 11}, 910 (2015).

\bibitem{Bayliss15} S. L. Bayliss, N. C. Greenham, R. H. Friend, H. Bouchiat, and A. D.
Chepelianskii, ``Spin-dependent recombination probed through the
dielectric polarizability'', Nat Comms {\bf 6}, 8534 (2015).

\bibitem{Boehme03} C. Boehme and K. Lips, ``Theory of time-domain measurement of spin-dependent
recombination with pulsed electrically detected magnetic
resonance'', Phys. Rev. B {\bf 68}, 245105 (2003).

\bibitem{McCameyPRL10} D. R. McCamey, K. J. van Schooten, W. J. Baker, S.-Y. Lee, S.-Y.
Paik, J. M. Lupton, and C. Boehme, ``Hyperfine-field-mediated spin
beating in electrostatically bound charge carrier pairs'', Phys.
Rev. Lett. {\bf 104}, 017601 (2010).

\bibitem{GR13} R. Glenn, W. J. Baker, C. Boehme, and M. E. Raikh, ''Analytical description of
spin-Rabi oscillation controlled electronic transitions rates
between weakly coupled pairs of paramagnetic states with
$S=1/2$'', Phys. Rev. B {\bf 87}, 155208 (2013).

\bibitem{Eickelkamp} T. Eickelkamp, S. Roth, and M. Mehring,``Electrically detected magnetic
resonance in photoexcited fullerenes'', Mol. Phys. {\bf 95}, 967
(1998).

\bibitem{LandLif} L. D. Landau and L. M. Lifshitz,
{\it Quantum Mechanics, Non-Relativistic Theory (Vol. 3)}
(Butterworth- Heinemann, Oxford, 1981).

\bibitem{Dicke} R. H. Dicke, ``Coherence in spontaneous radiation processes'',
Phys. Rev. {\bf 93}, 99 (1954).

\bibitem{Flatte} N. J. Harmon and M. E. Flatt{\'e},
Phys. Rev. B {\bf 85}, 075204 (2012).

\bibitem{Slichter} C. P. Slichter, {\it Principles of Magnetic Resonance} (Harper \& Row,
New York, 1963).

\bibitem{DaneNatMat08} D. R. McCamey, H. A. Seipel, S.-Y. Paik, M. J. Walter, N. J.
Borys, J. M. Lupton, and C. Boehme, ``Spin Rabi flopping in the
photocurrent of a polymer light-emitting diode'', Nat. Mater. {\bf
7}, 723 (2008).

\bibitem{Desai07} P. Desai, P. Shakya, T. Kreouzis, and W. P. Gillin, ``Magnetoresistance in
organic light-emitting diode structures under illumination'',
Phys. Rev. B {\bf 76}, 235202 (2007).

\bibitem{Drew10} J. Y. Song, N. Stingelin, A. J. Drew, T. Kreouzis, and
W. P. Gillin, ``Effect of excited states and applied magnetic
fields on the measured hole mobility in an organic
semiconductor'', Phys. Rev. B {\bf 82}, 085205 (2010).

\bibitem{Koopmans11} A. J. Schellekens, W. Wagemans, S. P. Kersten, P. A. Bobbert,
and B. Koopmans, ``Microscopic modeling of magnetic-field effects
on charge transport in organic semiconductors'', Phys. Rev. B {\bf
84}, 075204 (2011).

\bibitem{Kipp15} K. J. van Schooten, D. L. Baird, M. E. Limes, J. M. Lupton, and C.
Boehme, ``Probing long-range carrier-pair spin–spin interactions
in a conjugated polymer by detuning of electrically detected spin
beating'', Nat. Commun. {\bf 6}, 6688 (2015).

\bibitem{List01} E. J. W. List, C.-H. Kim, A. K. Naik, U. Scherf, G. Leising, W.
Graupner, and J. Shinar, ``Interaction of singlet excitons with
polarons in wide band-gap organic semiconductors: A quantitative
study'', Phys. Rev. B {\bf 64}, 155204 (2001).

\end{references}
\end{document}